\documentclass[natbib]{emulateapj}
\input epsf

\begin{document}
\title{Two regimes of Turbulent Fragmentation and the stellar IMF  from Primordial to Present Day Star Formation}

\author{Paolo Padoan\altaffilmark{1}, \AA ke Nordlund\altaffilmark{2}, 
Alexei G. Kritsuk\altaffilmark{1}, Michael L. Norman\altaffilmark{1},
Pak Shing Li} \altaffilmark{3}

\altaffiltext{1}{Department of Physics, University of California, San Diego, 
CASS/UCSD 0424, 9500 Gilman Drive, La Jolla, CA 92093-0424; ppadoan@ucsd.edu}
\altaffiltext{2}{Astronomical Observatory / NBIfAFG, Juliane Maries Vej 30, DK-2100, Copenhagen, Denmark}
\altaffiltext{3}{Astronomy Department, University of California, Berkeley, CA 94720}

\begin{abstract}

The Padoan and Nordlund model of the stellar initial mass function (IMF) is 
derived from low order statistics of supersonic turbulence, neglecting gravity
(e.g. gravitational fragmentation, accretion and merging). In this work the predictions 
of that model are tested using the largest numerical experiments of supersonic hydrodynamic
(HD) and magneto-hydrodynamic (MHD) turbulence to date ($\sim 1000^3$ 
computational zones) and three different codes (Enzo, Zeus and the Stagger
Code). The model predicts a power law distribution for large masses, related 
to the turbulence energy power spectrum slope, and the shock jump conditions. 
This power law mass distribution is confirmed by the numerical experiments. 
The model also predicts a sharp difference between the HD and MHD regimes, 
which is recovered in the experiments as well, implying that the magnetic 
field, even below energy equipartition on the large scale, is a crucial 
component of the process of turbulent fragmentation. These results suggest
that the stellar IMF of primordial stars may differ from that in later epochs of star
formation, due to differences in both gas temperature and magnetic field strength. 
In particular, we find that the IMF of primordial stars born in turbulent clouds 
may be narrowly peaked around a mass of order 10~M$_{\odot}$, as long
as the column density of such clouds is not much in excess of $10^{22}$~cm$^{-2}$.

\end{abstract}


\section{Introduction}

In the turbulent fragmentation model of Padoan \& Nordlund (2002),
the mass distribution of gravitationally unstable cores in turbulent clouds 
is derived from low order statistics of supersonic turbulence  (the one-point 
pdf of gas density, and the first-order scaling of velocity differences -though 
we refer to the velocity power spectrum, as a second order proxy of 
the first order scaling) and from focusing on the fundamental flow geometry of 
shock compressions. This is a natural (if not traditional) approach to the solution of 
a very complex turbulence problem, because a simple solution must be based on low
order statistics giving the basic variance and scaling of the process. But
low order statistics cannot capture the flow geometry. Instead of relying
on closure models of a statistical nature, an alternative is to 
impose some knowledge of a fundamental flow feature that would
otherwise be hard to capture with statistical quantities. This is especially true
because of the strong intermittent nature of turbulent flows, meaning that
the most important flow structures could be properly accounted for only
by very high order statistics, so relatively low order statistical closure models 
are bound to largely overlook those important structures. A phenomenological model 
centered on the geometry of those structures, in combination with low order statistics is
therefore a valid alternative, and it is the approach of choice in the work
of Padoan \& Nordlund (2002). 

The model assumes that: 
i) the turbulence has a power law velocity power spectrum (the model really 
uses the first order scaling of velocity differences, but that is related 
to the power spectrum anyway); ii) cores
are formed by shocks in the turbulent flow and have size and 
density that scale as the postshock layer thickness and density;
iii) the number of such shocks scales self-similarly as the 
inverse of the cube of their size; iv) the condition for the 
collapse of small cores is that their mass exceeds their 
Bonnor-Ebert mass, derived from the lognormal probability
density function (pdf) of the gas density independently of the 
core mass. 

These assumptions are all reasonable for an approximately isothermal 
and supersonic turbulent gas. 
The first assumption is a well established result for turbulent flows. The second 
assumption is suggested by the simulations, where the unstable cores
are always found to be the densest regions of postshock sheets or filaments.
The third assumption is motivated by the very large Reynolds number of the
turbulence in star-forming clouds,  which is expected to generate a very extended
inertial range of scales and possibly a self-similar flow responsible for the network
of shocks (this self-similarity does not imply necessarily a hierarchical, space-filling
nested structure, where smaller compressing regions would always be inside larger ones).
Finally, the fourth assumption stems from the fact that most of the densest gas is 
located within dense cores, so the high density tail of the gas density pdf 
should be well represented by the distribution of the core mean density.

These assumptions result in a power law mass
distribution reflecting the scale-free nature of the turbulence,
and a turnover at small masses, where gas pressure competes 
with self-gravity. After integrating over the probability of 
exceeding the Bonnor-Ebert mass, the mass distribution is given by
the following formula:
\begin{figure}[ht]
%\centerline{
\epsfxsize=8.6cm \epsfbox{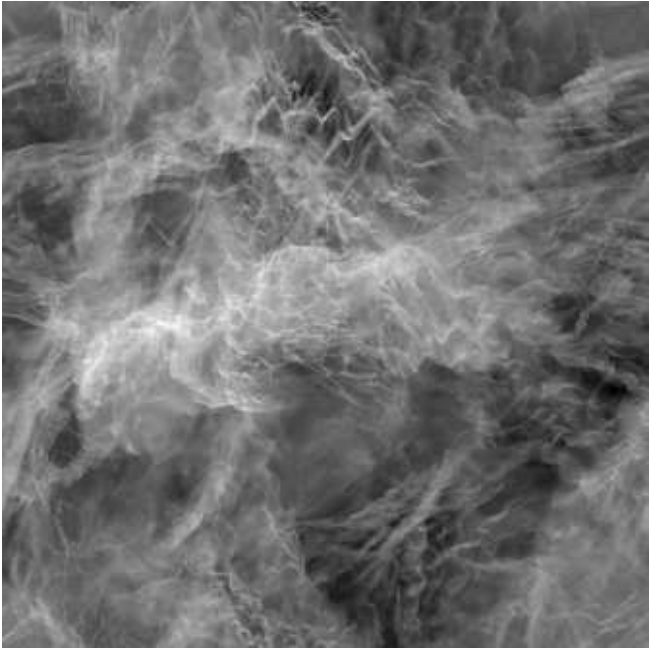}
%}
\caption[]{Logarithm of projected density from a snapshot of the 
Stagger-Code HD run.
}
\label{f1}
\end{figure}
\begin{equation}
N(m)dm= C\left[1+{\rm erf}\left({4{\rm ln}(m)+\sigma^2}\over{2 \sqrt{2}\sigma}\right)\right] m^{-x}dm
\label{imf}
\end{equation}
where the mass $m$ is in units of the average Bonnor-Ebert mass,
\begin{equation}
m_{\rm BE,o}=3.3 {\rm M}_{\odot}\left(n_{\rm o} \over{10^3{\rm cm}^{-3}}\right)^{-1/2}
\left(T_{\rm o} \over{10{\rm K}}\right)^{3/2},   
\label{be}
\end{equation}
$\sigma$ is the standard deviation of the gas density pdf (assumed to
be a lognormal) related to the rms Mach number of the turbulence (the 
sonic or the Alfv\'{e}nic Mach number in the HD or MHD regime respectively, 
see below):
\begin{equation}
\sigma= \sqrt{{\rm ln}(1+M_{\rm o}^2/4)}
\label{sigma}
\end{equation}
(from here on the subscript $\rm o$ denotes quantities averaged on 
the outer scale of the turbulence).
The coefficient $C$ depends on the physical parameters and is not 
discussed here because the normalization of the mass distribution 
is not required for this work. The power law slope, $x$, is determined 
by the power law slope of the energy spectrum, $\beta$ 
($\beta\approx 5/3$ in incompressible turbulence and $\beta=2$ 
in Burgers zero-pressure model), and by the shock jump conditions: 
\begin{equation}
x= 3/(4-\beta)
\label{x1}
\end{equation}
for $B \ge B_{\rm cr}$ (MHD jump conditions), and 
\begin{equation}
x=3/(5-2\beta)
\label{x2}
\end{equation}
for $B < B_{\rm cr}$ (isothermal HD jump 
conditions). The critical magnetic field value that separates 
the two regimes is given by the condition that the postshock gas 
pressure is of order the postshock magnetic pressure, corresponding
to an rms Alfv\'{e}nic Mach number, $M_{\rm A}$, of the order of 
the ratio of the mean gas and magnetic pressures, 
$M_{\rm A}\sim P_{\rm g}/P_{\rm m}$. This condition gives
\begin{equation}
B_{\rm cr}\approx 2\,\mu{\rm G}\left(T_{\rm o} \over{10{\rm K}}\right)
\left(u_{\rm o}\over{1{\rm km/s}} \right)^{-1}\left(n_{\rm o}\over{10^3{\rm cm}^{-3}}\right)^{1/2}
\label{bcr}
\end{equation}
\begin{figure}[ht]
%\centerline{
\epsfxsize=8.6cm \epsfbox{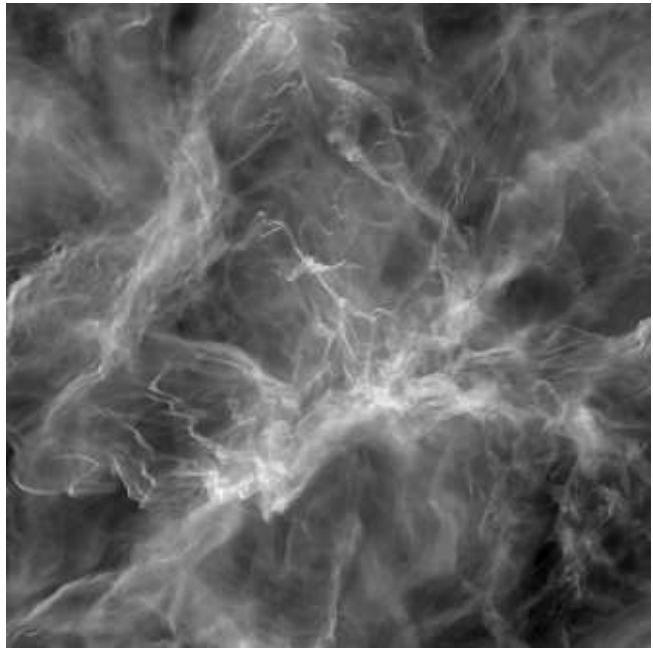}
%}
\caption[]{Logarithm of projected density from a snapshot of the 
Stagger-Code MHD run.
}
\label{f2}
\end{figure}
Because the Galactic magnetic field strength is locally $6\pm 2$~$\mu$G 
(Beck 2001; Han, Ferri\`{e}re, \& Manchester 2004), and perhaps larger 
in molecular cloud cores (Crutcher 1999; Bourke et al. 2001), current
star formation in the galactic disk occurs under MHD conditions. For a
value of $\beta=1.9$, found from the least dissipative simulations in this work, the predicted slope
of the mass distribution of prestellar cores is then $x= 1.4$, practically
the same as the Salpeter slope of the stellar IMF ($x=1.35$, Salpeter 1955). For 
very weak magnetic fields, perhaps in protogalaxies at very large redshifts, the slope
would be $x=2.5$, assuming again $\beta=1.9$. Furthermore, conditions at
high redshifts and very low or zero metallicity would also include a larger temperature, 
$T>100$~K (e.g. Palla, Salpeter, \& Stahler 1983; Abel, Bryan \& Norman 2000). 
The larger temperature results in a value of $B_{\rm cr}$ at least 
10 times larger than in present day environments with the same rms velocity 
and mean density, making the HD regime, and hence the steeper IMF, even more
likely to occur for stellar populations at high redshift. At the same time, the 
larger temperature also shifts the peak 
of the IMF to larger masses, according to equations (\ref{imf}) and (\ref{be}).
Although the first population III stars are usually believed to form in isolation
in the cores of very small halos,  population III stars formed somewhat later
and in somewhat more massive halos (Jimenez \& Haiman 2006; Iliev et al. 2006) 
may indeed emerge from turbulent star-forming environments more akin to current star 
formation sites. Thus, the steeper IMF for massive stars corresponding to the HD 
regime may also apply to bona-fide population III stars, as well as to early 
population II stars.

The peak of the distribution shifts to smaller masses with increasing 
Mach number and gas density, which also increases the abundance of
low mass stars and brown dwarfs. 
\begin{figure}[ht]
\centerline{
\epsfxsize=8.6cm \epsfbox{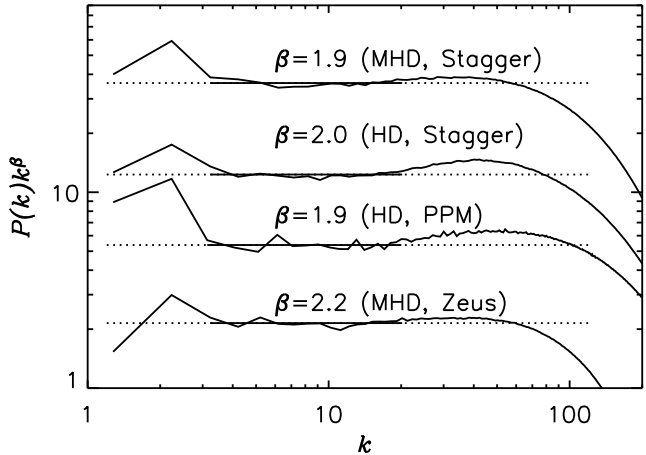}
}
\caption[]{Compensated power spectra of the main four runs and 
least square fits in the range of wavenumbers $3\le k \le 20$.
}
\label{f3}
\end{figure}
For reasonable values of the physical 
parameters, this mass distribution becomes essentially the same as the 
observed stellar IMF in Chabrier (2003). This coincidence suggests that
the process of turbulent fragmentation may play a major role in the
origin of the stellar IMF (Padoan \& Nordlund 2002), with only minor 
effects due to gravitational fragmentation, accretion or merging.

A numerical estimate of the dependence of the peak mass on 
the rms Mach number gives:
\begin{equation}
m_{\rm p}=3.0 M_0^{-1.1} M_{\rm BE,0}
\label{mp1}
\end{equation}
assuming $x=1.4$ (MHD regime), and 
\begin{equation}
m_{\rm p}=3.9 M_0^{-1.7} M_{\rm BE,0}
\label{mp2}
\end{equation}
assuming $x=2.5$ (HD regime). In the following we relate $m_{\rm p}$ to the
peak of the stellar IMF. Since the latter can be accurately determined observationally 
only for stellar clusters, presumably formed in gravitationally bound cores, we may 
assume the gas velocity dispersion is of order the virial velocity. With this assumption,
and using the rms Alfv\'{e}nic Mach number, equation (\ref{mp1}) gives:
\begin{equation}
m_{\rm p}=0.25 {\rm M}_{\odot} N_{0, 22} ^{-1.1}B_{0,10} ^{1.1} n_{0,4}^{-1/2}T_{0,10}^{3/2}
\label{mp1b}
\end{equation}
where $N_{0,22}$, $B_{0,10}$, $n_{0,4}$, and $T_{0,10}$ are the mean 
column density, magnetic field, particle density and temperature in units
of $10^{22}$~cm$^{-2}$, 10~$\mu$G, $10^4$~cm$^{-3}$, and 10~K
respectively, which are typical values for star-forming cloud cores. This value
of $m_{\rm p}$ provides an estimate of the peak of the stellar IMF (although it 
should be somewhat larger than the IMF peak because a fraction of the core 
mass is not accreted onto the star and because some cores may result in binaries or 
multiple systems). 

The observed IMF (of multiple systems) peaks at roughly 0.2~M$_{\odot}$ 
(Chabrier 2003), with no clear 
evidence of a strong departure from this value in any star-forming region, with the 
exception of the Taurus molecular cloud complex (Luhman 2004). As both temperature 
and column density don't have very large variations from cloud to cloud, our fragmentation 
model would predict a nearly constant value of $m_{\rm p}$ if $B_0\propto n_0^{0.45}$, for which
there is some observational and theoretical support (Myers \& Goodman 1988; Crutcher 1999; 
Padoan \& Nordlund 1999; Basu 2000). 
\begin{figure}[ht]
\centerline{
\epsfxsize=8.6cm \epsfbox{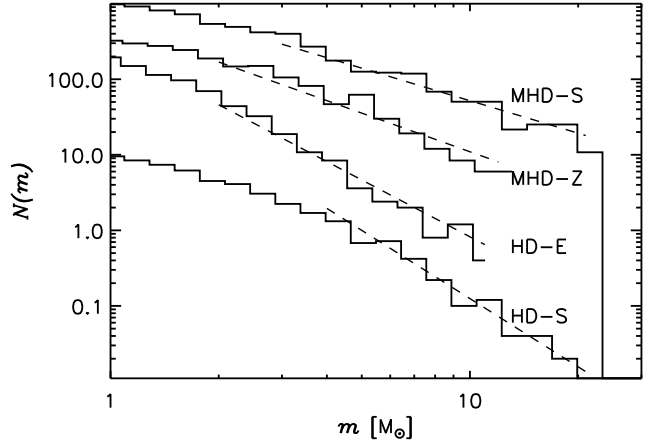}
}
\caption[]{Mass distributions of gravitationally unstable cores above 
1~M$_{\odot}$, for the main four experiments scaled to a mean density 
of $10^4$~cm$^{-3}$, a box size of 6~pc, and a clumpfind density 
resolution $f=8$\%. The dashed lines show the power law derived 
from the power spectrum slope and the shock jump conditions of the
corresponding simulations, according to the turbulent fragmentation model.
The histograms are arbitrarily offset in the vertical direction for clarity.
}
\label{f4}
\end{figure}
Sites of massive star formation may have larger mean 
column density, but they also have larger temperature than regions of low-mass star 
formation (e.g. Ossenkopf, Trojan, \& Stutzki 2001; Beuther et al. 2006), which may 
result in a similar value of $m_{\rm p}$.   

In the HD regime, assuming again virial velocity dispersion and using the sonic
rms Mach number, the peak mass given by equation (\ref{mp2}) is:
\begin{equation}
m_{\rm p}=0.16 {\rm M}_{\odot} N_{0, 22} ^{-1.7} n_{0,4}^{0.35}T_{0,10}^{2.35}
\label{mp2b}
\end{equation}
Assuming for example that the earliest turbulent conditions are found within halos 
of approximately $10^8$~M$_{\odot}$, cooled to the background temperature by 
non-equilibrium formation of HD (Johnson \& Bromm 2006), approximate values of 
mean density, temperature, and column density are $n_0=1$~cm$^{-3}$, $T_0=43$~K, 
and $N_0=10^{21}$~cm$^{-2}$, giving $m_{\rm p}=10$~M$_{\odot}$.
It has been often proposed that the IMF of population III stars should contain
an excess of massive stars relative to the present day IMF. This may indeed occur
with our IMF in the HD regime, despite its steep slope, due the high gas temperature
of primordial gas, shifting the peak of the IMF toward large masses. Because of its 
steep slope, though, {\it the IMF of primordial stars would then be narrowly 
peaked around a large mass of order} $10$~M$_{\odot}$.  On the other hand, a larger 
column density of order $10^{22}$~cm$^{-2}$ would give $m_{\rm p}=0.2$~M$_{\odot}$, 
similar to the IMF peak in regions of present-day star formation. This example shows that 
the typical mass of primordial stars born in turbulent clouds cannot be estimated without  
a reliable value of the average column density of such clouds.

\section{The Simulations}

As the turbulent fragmentation model outlined in the previous section neglects 
gravity (apart from the selection of unstable cores), its predictions can be 
tested with numerical simulations of
supersonic MHD turbulence. However, because the model relies on the 
scale-free nature of the turbulence, it cannot be easily tested unless
\begin{figure}[ht]
\centerline{
\epsfxsize=8.6cm \epsfbox{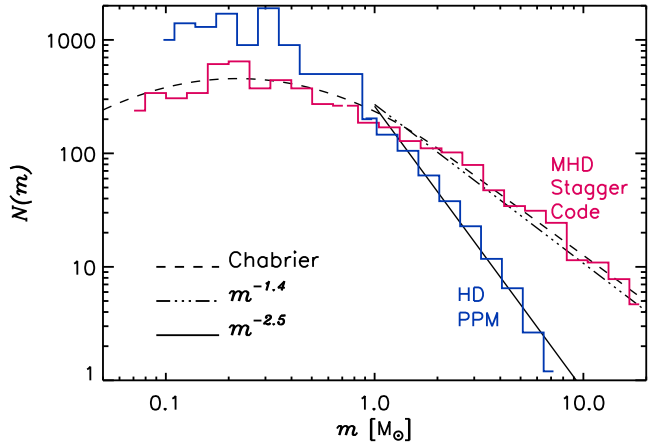}
}
\caption[]{Mass distributions of gravitationally unstable cores from the 
HD and MHD regimes (Enzo and Stagger Code respectively), computed with 
$f=16$\% and assuming a mean gas density of $10^4$~cm$^{-3}$. Each mass 
distribution is the result of matching two mass distributions, computed 
for a computational box size of 1~pc and 6~pc. The Chabrier IMF (Chabrier 2003)
and the fragmentation model predictions for the power law mass distributions
of each run are also plotted. 
}
\label{f5}
\end{figure}
an extended inertial range of the turbulence is generated in the simulations,
which requires both large numerical resolution (large computers) and low 
numerical diffusivity (good codes). The main comparison between the MHD and
HD regimes is here based on simulations with the Stagger Code, on a numerical
mesh of $1,000^3$ computational zones. In order to rule out numerical artifacts
of a specific code, we also simulated the HD regime with the Enzo code 
(Norman \& Bryan 1999), and the MHD regime with the Zeus code (Stone \& 
Norman 1992), in both cases on a numerical mesh of $1,024^3$ 
computational zones. The three codes are quite different from each other.
Enzo is based on a Rieman solver and PPM reconstruction, the Stagger code is 
based on a high order (5th order in space, 4th order in time) finite different scheme, 
and Zeus is a low order (2nd order in space, 1st order in time) finite difference code. 

All simulations make use of periodic boundary conditions, isothermal
equation of state, random forcing in Fourier space at wavenumbers 
$1\le k\le 2$ ($k=1$ correspond to the computational box size), uniform 
initial density and magnetic field (in the MHD runs), random 
initial velocity field with power only at wavenumbers $1 \le k\le 2$.
The simulations are run for several dynamical times to relax the
turbulence at rms Mach numbers of 6 or 10, before being analyzed. 
All results in this paper are for rms Mach number 10, unless otherwise 
specified . In Figures~1 and 2 we show two projections
of the density field, from the HD and MHD Stagger Code runs. 
The density field in the HD run appears to be
significantly more fragmented than its MHD counterpart. There are two 
main reasons for this difference: i) The density contrast in the HD
shocks is larger than in the MHD shocks, creating thinner postshock 
layers from shocks with equal sonic Mach number; ii) the HD postshock
layers are Kelvin-Helmholtz unstable, due to the strong shear flow that
originates in oblique shocks, while in most of the MHD layers
the same instability is suppressed by the magnetic field that is
\begin{figure}[ht]
\centerline{
\epsfxsize=8.6cm \epsfbox{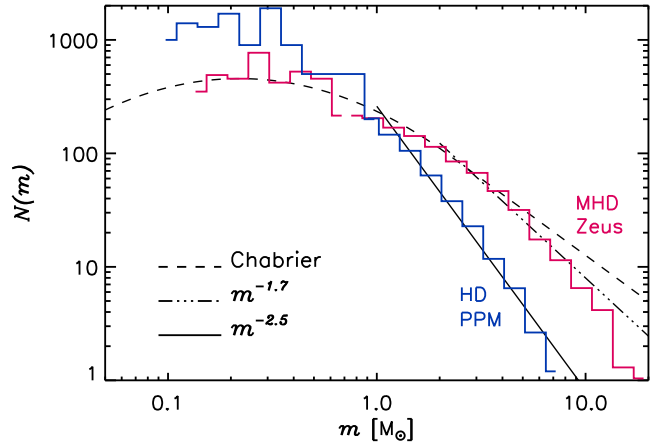}
}
\caption[]{Same as Figure~5, but for the Zeus MHD run instead of the  
Stagger-Code MHD run. Notice how the steeper Zeus mass distribution
is recovered. The model predicts the Zeus mass distribution to be
steeper than the Stagger-Code one, as a result of the steeper turbulence 
power spectrum in the more diffusive Zeus run.
}
\label{f6}
\end{figure}
amplified in the compression. The turbulent fragmentation model
of Padoan and Nordlund (2002) makes explicit use of the 
shock jump conditions, which results in the prediction of a steeper mass 
distribution in the power law range of masses in the HD case than in 
the MHD one, but no direct reference to instabilities in the layers. 

Figure~3 shows the compensated power spectra of the four main simulations. 
The power spectra are defined as the squared of the modulus of the Fourier
transform of the velocity, integrated over a wave-number shell. If $\hat{\bf u}_{ i}(\bf k)$
is the Fourier transform of the $i$ velocity component, ${\bf u}_{i}({\bf r})$, the power
spectrum is $P_{i}(k)=\sum \hat{\bf u}_{ i}\hat{\bf u}^*_{i}$, where the sum is over 
all $i$ and all wave-numbers ${\bf k}$ in the shell $k \le |{\bf k}| < k+dk$. 
$P(k)$ is proportional to the contribution to the mean square velocity from 
all wave-numbers in the shell $k \le |{\bf k}| < k+dk$.

The plots in Figure~3 have been arbitrarily shifted in the vertical direction. 
Deviations of more than a factor of two from a power law fit are found 
only at wavenumbers $k>100$, so the turbulence is roughly scale-free
for almost two orders of magnitude in wavenumbers. The power law slopes
depend somewhat on the exact range of wavenumbers used in the fit.
We have chosen to measure the power spectrum slope in the range 
$3\le k\le 20$, because larger wavenumbers are affected by the bottleneck
effect (e.g. Falkovich 1994; Dobler et al. 2003; Haugen \& Brandenburg 2004). 
If the least square fits were extended up to $k=100$, to estimate
an effective power spectrum slope relevant for the turbulent fragmentation 
process, the slopes would be only slightly different. We get $\beta=1.9$ 
and 2.0 from the Stagger code in the MHD and HD regimes respectively. 
The Enzo code in the HD regime gives $\beta=1.9$, and Zeus in the MHD 
regime $\beta=2.2$. The corresponding values of the exponent of the power 
law range of the mass distribution of unstable cores are, according to the 
model of turbulent fragmentation, $x=1.4$ and 3 for the MHD and HD regimes 
of the Stagger code, and $x=1.7$ and 2.5 for the MHD and HD regimes of Zeus 
and Enzo respectively.

\begin{figure}[ht]
\centerline{
\epsfxsize=8.6cm \epsfbox{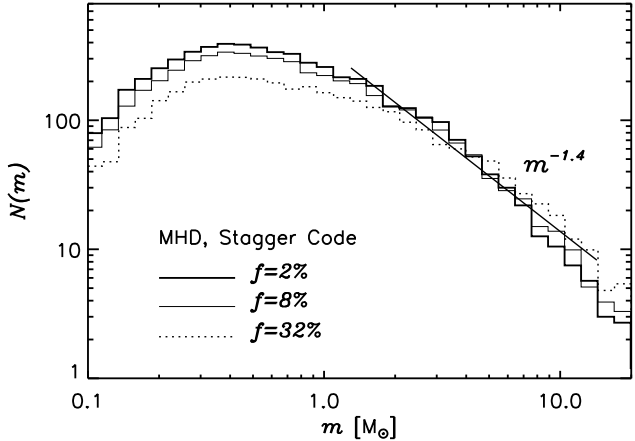}
}
\caption[]{Test of convergence of the mass distribution with decreasing value
of the density resolution parameter, $f$, from the Stagger-Code MHD run. 
The mass distribution is well converged at $f=8$\%.
}
\label{f7}
\end{figure}

The power spectra show that the Stagger code is only slightly more dissipative
than Enzo, while Zeus is significantly more dissipative than both. The numerical 
dissipation results in a power spectrum that is well fit by an extended power 
law, partly because an effect of the numerical dissipation is to suppress the 
bump of excess power corresponding to the bottleneck effect, visible especially 
in the least dissipative code, Enzo, around $k=40$. The power law form of the
power spectrum may therefore be deceiving (as evidence of numerical 
convergence), because it's slope is apparently
dependent on the numerical diffusivity. It is only by comparing different codes
or the same code with different values of the diffusivities that we can 
find a truly converged power spectrum. If the power spectrum is not converged 
to its correct slope, due to a lack of dynamic range of scales or an excess of
numerical diffusivity, the slope of the mass distribution may be strongly 
affected. For example, in the MHD case we get $x=1.4$ from the Stagger code,
and $x=1.7$ from the Zeus code. It is possible that even the Stagger code power 
spectrum is not fully converged, and that the correct value is approximately
$\beta=1.8$, in which case the slope is $x=1.36$. This is suggested by the
fact that in the HD regime PPM has a slightly smaller exponent than in 
the HD regime of the Stagger code. Padoan et al. (2006) have recently 
obtained an accurate measurement of the velocity power spectrum in the 
Perseus molecular cloud complex.  Their result is $\beta=1.81\pm 0.10$,
accurate enough to rule out the significantly larger power spectrum slopes
produced by more dissipative SPH simulations (see \S~4).

\begin{figure}[ht]
\centerline{
\epsfxsize=8.6cm \epsfbox{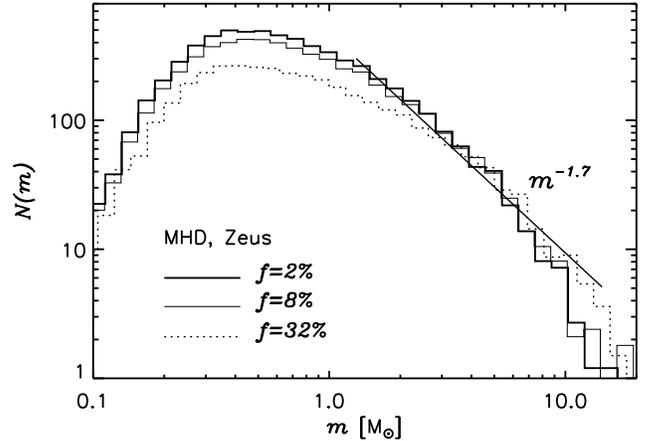}
}
\caption[]{Same as Figure~7, but from the Zeus MHD run.
}
\label{f8}
\end{figure}

\section{Mass Distributions}

We compute the mass distribution of gravitationally unstable cores 
formed in turbulence simulations without self-gravity primarily
to learn about the effect of turbulence under different conditions,
for example with and without magnetic fields, and to compare with
the predictions of the turbulent fragmentation model. The question
of how self-gravity would modify the mass distribution is a separate
one, and will not be addressed in this work. However, it is important
to stress that the present result are obtained after the driven turbulence 
has been statistically relaxed, which could not be achieved with self-gravity.
The mass distributions derived in this work and the mass distribution
predicted by Padoan \& Nordlund (2002), can therefore represent, at best,
a guess of the final outcome of more realistic simulations with self-gravity.
In such simulations including self-gravity the mass distribution of unstable cores
may initially vary with time, as the most massive cores are still being assembled
by converging turbulent flows while their central part has already collapsed.

Cores are defined as connected overdensities that cannot be split 
into two or more overdensities of amplitude $\delta \rho/\rho>f$.
The unstable cores are simply the cores with mass larger than their 
Bonnor-Ebert mass. 
These definitions are implemented  
in our clumpfind algorithm by scanning the density field with
discrete density levels, each of amplitude $f$ relative to the 
previous one. Only the connected regions above each density level
that are larger than their Bonnor-Ebert mass are selected as unstable
cores. After this selection, the unstable cores from all levels form 
a hierarchy tree. Only the final (unsplit) core of each branch is retained.
It is important to impose the Bonnor-Ebert condition while building the 
tree, otherwise some large unstable cores would be incorrectly 
eliminated if they split into smaller cores that were later rejected 
based on the Bonnor-Ebert condition.

Clumpfind algorithms differ in the way they assign the surrounding 
mass to the cores. The popular algorithm by Williams, de Geus, 
\& Blitz (1994), for example, uses up all the available mass (see their
Figure~2). This results in a core formation efficiency of 100\% above the 
threshold density, which is an artifact of that algorithm with no physical 
justification, though it may mimic  a process of competitive accretion. 
Our algorithm, instead,
assigns to each core only the mass within the density isosurface
that defines the core (below that density level the core would be
merged with its next neighbor). We prefer this definition because
the smallest possible mass is assigned to each core and we want
to study the effect of turbulent fragmentation in isolation, 
not trying to guess the outcome of the subsequent accretion.
It turns out that, with this definition, and under conditions typical
of molecular clouds, the unstable cores contain a few percent of the 
total mass. 
\begin{figure}[ht]
\centerline{
\epsfxsize=8.6cm \epsfbox{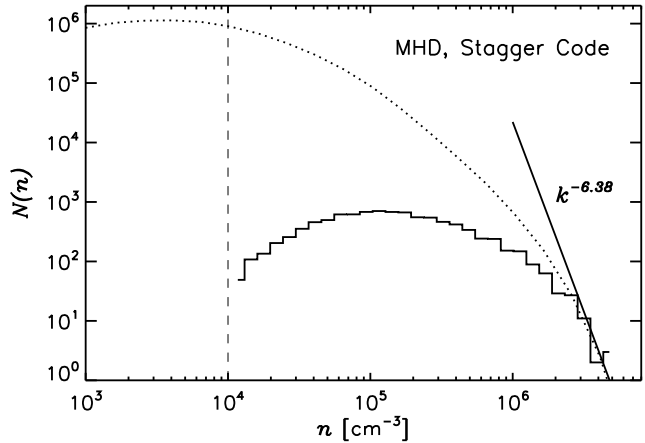}
}
\caption[]{Mean density distributions of the unstable cores selected from
two snapshots of the Stagger Code MHD experiment (solid histogram), assuming 
a mean gas density of $10^4$~cm$^{-3}$ and a 6~pc size, as in previous figures. The dotted 
curve is the pdf of gas density of the whole computational domain, computed for the 
same two snapshots. The solid straight line shows the power law distribution of
core mean densities that would be derived from the velocity scaling, 
$N(n)\propto n^{-6/(\beta-1)}$, using the specific value of $\beta=1.94$ found in those
two snapshots. The vertical dashed line marks the mean gas density. The two curves 
and the power law are normalized to approximately match each other at the largest 
densities, to illustrate that a very large fraction of the densest gas is found within 
unstable cores, while gas of decreasing density is increasingly harder to capture in
unstable regions.  
}
\label{f9}
\end{figure}
This suggests that not much of the remaining mass will 
ever be accreted, as we know that the star formation efficiency in
molecular clouds is approximately a few percent. 

Once the physical size and mean density of the system are chosen,
the clumpfind algorithm depends only on two parameters: i) The 
spacing of the discrete density levels, $f$, and ii) the minimum
density above which cores are selected, $\rho_{\rm min}$. In principle
there is no need to define a minimum density, but in practice it
speeds up the algorithm. We have verified that results (including the
total mass in cores) do not change 
significantly for values of $\rho_{\rm min}$ below the mean gas density,
so we scan the density field only above the mean density. Notice
that only half of the volume, but most of the mass, is found above the mean 
density, because, according to the lognormal pdf, most of the mass is packed
in a small volume fraction. 

The parameter $f$ may be chosen according to
a physical model providing the value of the smallest density fluctuation
that could collapse separately from its contracting background. However,
given the difficulty of predicting the outcome of the gravitational 
fragmentation, we prefer to simply search for a convergence of the mass
distribution with decreasing values of $f$. Luckily, the convergence is
typically obtained already at a value of $f\approx 16$\%, meaning that
differences between the mass distributions with $f=8$\% and 16\% are generally 
insignificant. The cores are therefore well defined, and in most cases
they correspond to density fluctuations, relative to the surrounding gas,
even much larger than $f$.

In Figure~4 the mass distributions above 1~M$_{\odot}$ are plotted for
the main four experiments scaled to a mean density of $10^4$~cm$^{-3}$,
a box size of 6~pc, and a clumpfind density resolution $f=8$\%. 
\begin{figure}[ht]
\centerline{
\epsfxsize=8.6cm \epsfbox{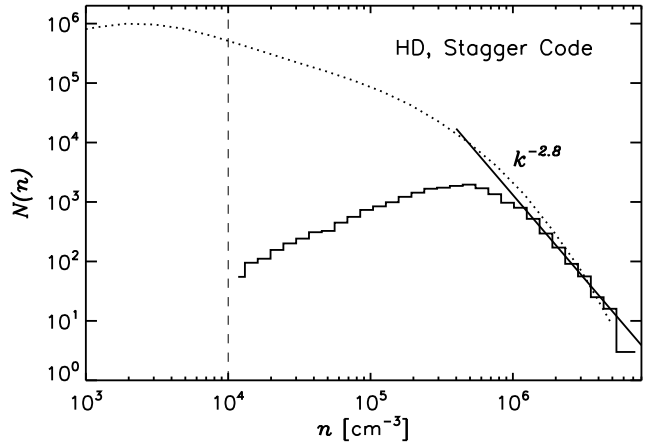}
}
\caption[]{Same as Figure~\ref{f9}, but for one snapshot of the 
Stagger Code HD run. The power law here is $N(n)\propto n^{-3/(\beta-1)}$, with the 
value of $\beta=2.06$ computed for this snapshot.
}
\label{f9b}
\end{figure}
Overplotted on the corresponding power law section of each mass distribution, the
dashed lines show the power law derived from the power spectrum slope and
the shock jump conditions of each simulation, according to the turbulent
fragmentation model, $x= 3/(4-\beta)$ in the MHD regime, and $x=3/(5-2\beta)$
in the HD regime. The general trend is recovered well, despite deviations to 
be expected because this mass distributions are from single snapshots, 
not time averages.

Figure~5 shows the mass distributions of the HD and MHD regimes (Enzo and
Stagger Code respectively), computed with $f=16$\% and assuming a mean
gas density of $10^4$~cm$^{-3}$. Each mass distribution is the result
of matching two mass distributions, computed for a computational box size 
of 1~pc and 6~pc. The 6~pc case makes it possible to sample masses in 
the range $1-10$~M$_{\odot}$, and hence to probe the effect of the turbulence
power spectrum and shock jump conditions on the mass distribution, but suffers
from incompleteness for stars below $1$~M$_{\odot}$. The 1~pc case samples 
well the turnover region, and hence defines the peak mass for that mean 
density and rms Mach number, but does not yield intermediate and high mass stars.
Figure~6 is equivalent to Figure~5, but uses the MHD Zeus run, instead of
the Stagger Code run.

The numerical mass distributions reproduce the sharp difference between the
HD and MHD regimes predicted by the turbulent fragmentation model. The steeper
mass distribution expected from the Zeus run, compared with the Stagger Code run,
due to the steeper Zeus turbulence power spectrum, is also recovered.
The slopes predicted by the turbulent fragmentation model are overplotted
in each figure. Furthermore, the MHD regime yields a mass distribution of 
gravitationally unstable cores practically indistinguishable from Chabrier's 
stellar IMF (Chabrier 2003), both in the Zeus and in the Stagger Code runs.
Finally, the Stagger Code HD run also yields a mass distribution consistent
with the model prediction (see for example Figure~4). The relation between 
the mass distribution and the power spectrum and shock jump conditions is
therefore successfully tested with 3 different codes at very high numerical
resolution. As discussed below, the model predictions for the HD regime are 
\begin{figure}[ht]
\centerline{
\epsfxsize=8.6cm \epsfbox{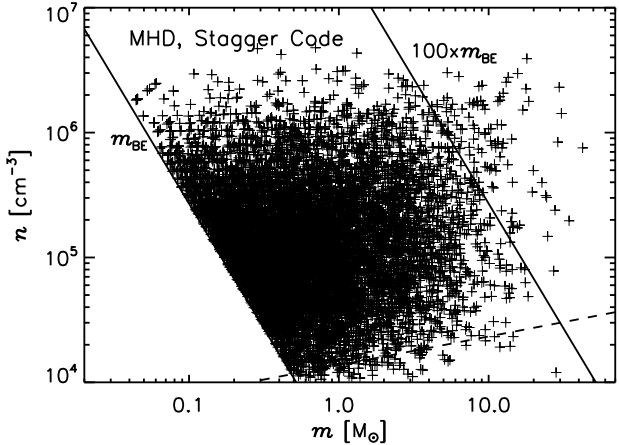}
}
\caption[]{Mean density versus mass, of the unstable cores selected 
from two snapshots of the Stagger Code MHD run, assuming a mean gas density of 
$10^4$~cm$^{-3}$ and a 6~pc size, as in previous figures. The solid line on the 
left marks the Bonnor-Ebert mass, while the solid line on the right a mass 100 times larger 
than the Bonnor-Ebert value. The turbulence produces a range of masses and mean
densities of unstable cores spanning three orders of magnitude. The dashed line is the 
power law relation between mean density and mass based on the average scale dependence 
of these two quantities, $n\propto m^{(\beta-1)/(8-2\beta)}$, using $\beta=1.94$, as in Figure~\ref{f9}. 
}
\label{f10}
\end{figure}
also qualitatively confirmed with a fourth code (the TVD code), at lower 
resolution.

To verify the convergence of the clumpfind algorithm with respect to the density
resolution, expressed by the parameter $f$, we plot in Figures~7 and 8 the mass
distributions of the MHD regime assuming a mean gas density of $10^4$~cm$^{-3}$
and a 6~pc size. Figure~7 is from the Stagger Code MHD run and Figure~8 from the
Zeus MHD run. For this convergence study we have computed together the mass 
distributions of unstable cores selected from three different snapshots. With 
these larger samples, statistical deviations are reduced, resulting in a
more sensitive test of convergence. Between $f=32$\% and $f=8$\% there is a 
tendency to fragment the largest cores and create a larger number of small 
cores. However, the differences between $f=8$\% and
 $f=2$\% are rather small. Furthemore, the slope of the mass distribution
above 2-3~M$_{\odot}$ is rather independent of resolution, even if the power 
law section of the mass distribution tend to move slightly to larger masses
at very large values of $f$.

\section{Discussion}

\subsection{Variance Versus Scaling}

The turbulent fragmentation model estimates the mass distribution of unstable
cores, based on the assumption that their size is determined by the thickness of
postshock layers. So far we have used the simulations and the clumpfind algorithm 
to confirm the predicted relation between the slope of the mass distribution and that
of the velocity power spectrum. The model also implies that, 
on the average, the {\it size} and {\it mean density} of cores should be scale dependent
as well, at least for cores significantly more massive than the average Bonnor-Ebert 
\begin{figure}[ht]
\centerline{
\epsfxsize=8.6cm \epsfbox{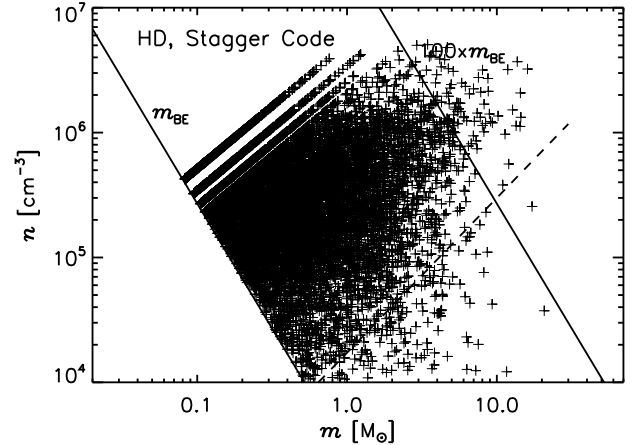}
}
\caption[]{Same as Figure~\ref{f10}, but for the cores selected from one snapshot
of the Stagger Code HD run. The dashed power law here is $n\propto m^{(\beta-1)/(5-2\beta)}$,
with the value of $\beta=2.06$ computed for this snapshot.
}
\label{f10b}
\end{figure}
mass. In the absence of variance, the distributions of core properties are derived from their scale 
dependence assuming the self-similar distribution of scales, $N(L)\propto L^{-3}$. A quantity 
$X$, scaling like $X\propto L^a$, would have a power law distribution $N(X)\propto X^{-3/a}$,
where the distributions are expressed as probability densities per logarithmic intervals 
(or, equivalently, they can be thought of as cumulative functions derived from the 
integration of probability densities per linear intervals). A very steep distribution of 
a quantity $X$ is therefore equivalent to a weak scale dependence of $X$ (small 
values of the exponent $a$), resulting in a limited range of values of $X$ caused by
its scale dependence alone. 

However, the properties of cores arising in turbulent supersonic flows are 
random variables following some distributions that can be roughly characterized 
by their standard deviation. This must be true on any scale. If the standard deviation 
of the distribution at a fixed scale is very large, it may exceed the range of values 
expected from the scale dependence, and the full distribution (integrated over all scales) 
may resemble more the distribution at a fixed scale than the power law predicted from the 
scale dependence. In the case of the core mass distributions in both the MHD and HD 
regimes, $a>1$, meaning that the logarithmic range of core masses exceeds the logarithmic 
range of scales, and therefore the scale dependence is expected to leave a strong imprint in the mass 
distributions. This is a case in which the scale dependence of a quantity results in a 
well defined power law distribution of that quantity. If instead the scale dependence
of a quantity $X$ is weak ($a<1$), it may be completely masked by the variance of 
its distribution at a fixed scale. 

Let's consider first the distribution of the core mean density. According to our
model, the scaling of velocity results in the core mean density distribution 
$N(n)\propto n^{-6/(\beta-1)}=n^{-6.67}$, in the
MHD regime, and $N(n)\propto n^{-3/(\beta-1)}=n^{-3.33}$ in the HD regime, where 
the numerical estimate of the exponents assumes $\beta=1.9$. These density distributions
are both steeper than the corresponding mass distributions, while the effect of the variance
in density should be comparable to that in mass, as mass is proportional to density.
Therefore, even if the mass distributions have power law tails, the core mean density 
\begin{figure}[ht]
\centerline{
\epsfxsize=8.6cm \epsfbox{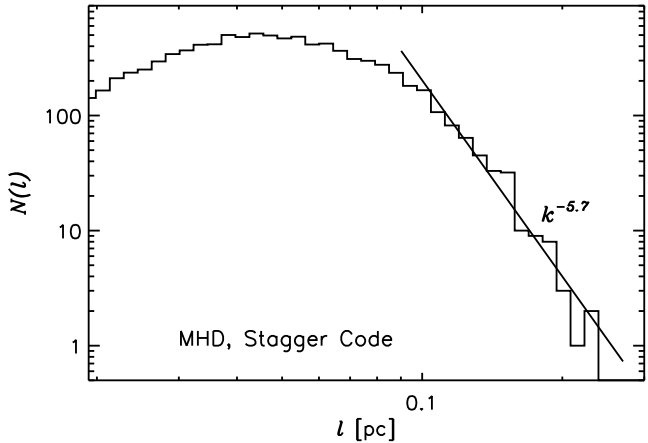}
}
\caption[]{Size distribution of the unstable cores selected from the same 
two snapshots of the Stagger Code MHD experiment as in the left panels of 
Figures~\ref{f9} and \ref{f10}. The core size is defined as the cubic root of its volume. 
The power law distribution resulting from the scale dependence of the core size, 
$N(l)\propto l^{-6/(3-\beta)}$, is plotted as a solid line, using the value of $\beta=1.94$ 
found in the two snapshots. 
}
\label{f11}
\end{figure}
distributions may not show any power law. This is certainly the case in the MHD
regime, given the very large exponent and hence the very weak scale dependence of the
mean density (very narrow range of densities over a large range of scales). 

In Figures~\ref{f9} and \ref{f9b} we show the MHD and HD mean density 
distributions of unstable cores from the Stagger Code experiments (solid histograms),
the power laws based on the velocity scaling (straight solid lines) and the pdf of gas
density from the whole computational volume (dotted curves). 
Consistent with the arguments above, in the
MHD regime there is no evidence of the power law scaling, while in the HD regime 
a short power law is visible, with a slope consistent with the model prediction. 
Notice how the core mean density distribution gradually departs from 
the general density pdf toward lower densities, as a result of the selection of only
gravitationally unstable cores. This large amount of gas mass not locked in unstable 
cores, even at relatively high densities, illustrates how the turbulence controls the
efficiency of the star formation process. 
 
The weak scale dependence of the core mean density can also be appreciated by
plotting the core mean density versus the core mass,  as done in Figures~\ref{f10} and \ref{f10b}.
These scatter plots don't show much evidence of the power law relation of mean
density and mass from the scale dependence of these two quantities,
$n\propto m^{(\beta-1)/(8-2\beta)}=m^{0.21}$ in MHD, and 
$n\propto m^{(\beta-1)/(5-2\beta)}=m^{0.75}$ in HD, assuming again $\beta=1.9$
in the numerical estimates. At any given mass, the density may span 
almost three orders of magnitude, although most values are within an factor
of ten range. This large variance is unavoidable in isothermal supersonic turbulence, 
due to the broad Lognormal density pdf, with standard deviation approximately
equal to half the rms Mach number of the turbulence.

\begin{figure}[ht]
\centerline{
\epsfxsize=8.6cm \epsfbox{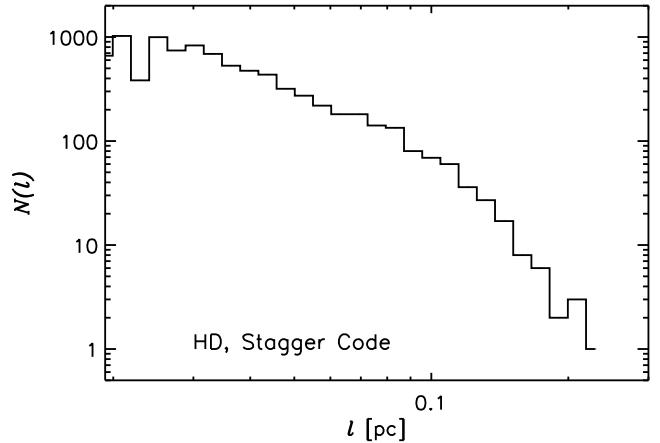}
}
\caption[]{Same as in Figure~\ref{f11}, but for the Stagger 
Code HD experiment. The scaling law would now give an extremely steep power law, 
$N(l)\propto l^{-3/(2-\beta)}$, or essentially no size range.  
}
\label{f11b}
\end{figure}

The average value of the core size should also be scale dependent, with the 
corresponding distributions being $N(l)\propto l^{-6/(3-\beta)}=l^{-5.45}$ in MHD
and  $N(l)\propto l^{-3/(2-\beta)}=l^{-30}$ in HD (assuming $\beta=1.9$ in the
numerical estimates). The MHD power law is rather steep, but the density
variance translates into a smaller size variance, as $l\propto n^{-1/3}$, so even
a rather weak scale dependence may result in a power law size distribution.
In the HD regime, there is almost a perfect cancellation of the scale dependence
of the shocked layer column density and volume density, making the average 
thickness of such a layer almost scale independent. The resulting power law distribution
would be so steep, that even a small variance would completely mask it. 

Figures~\ref{f11} and \ref{f11b} show the core size distribution of the same cores as in 
Figures~\ref{f9} to \ref{f10b}. The core size is defined as the cubic root of its 
volume. The power law distribution of the mean core size resulting from the
scale dependence is plotted as a solid line in the MHD case, with the exponent
corresponding to a value of $\beta=1.94$, as measured from the two snapshots
used to select the cores. In the MHD experiment there is some evidence
of a power law tail, with the same slope as inferred from the scale dependence
of our model. In the HD regime, instead, there is no evidence of the power law tail, as that
would correspond to a sharp cutoff, which is easily masked by the variance of the size 
distribution at a fixed scale. 

In summary, based on the competition between variance and scaling in establishing 
statistical distributions of core properties, we expect to find power law tails in the
distributions of masses and sizes in the MHD regime, and only masses (or marginally densities)
in the HD regime. The unstable cores selected from our numerical experiments have clear 
power law tails only in the case where these are expected, and in such cases the
power law slopes are those predicted by the model. In Padoan and Nordlund
(2002) the competition between the variance and the scaling is responsible
for the flattening and the turnover of the mass distribution toward smaller masses,
because toward smaller masses the selection of the unstable cores becomes 
increasingly sensitive to their density. At large masses, because the scaling is 
the dominant effect, the variance is neglected. 

\begin{figure}[ht]
\centerline{
\epsfxsize=8.6cm \epsfbox{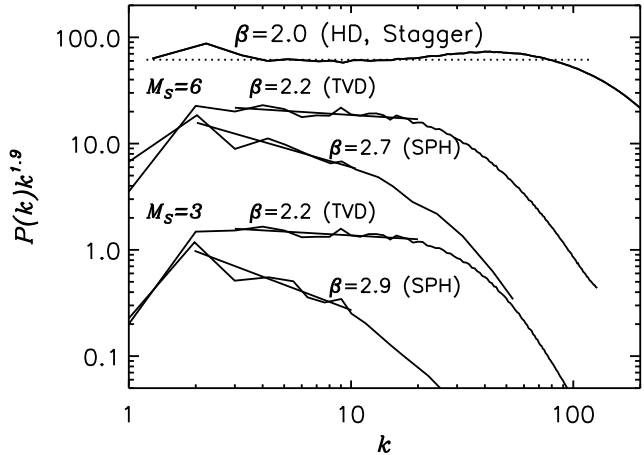}
}
\caption[]{Power spectra compensated for the slope of the Stagger-Code HD 
run, $\beta=1.9$. The TVD and SPH power spectra are the same as in Figure~2
of Ballesteros-Paredes et al. (2006), for the Mach numbers 3 and 6.
}
\label{f12}
\end{figure}

In numerical simulations,  even when the variance should not be dominant, power 
law tails may be absent from statistical distributions of core properties, as a result of 
the limited range of scales relative to the actual interstellar turbulence.
If the range of scales is reduced, the range of values in core properties due to the scale
dependence is also reduced, possibly to the point of becoming smaller than
the variance of the distribution at a fixed scale. Lognormal-like tails may then 
cut short the power law distributions, as a numerical effect.

It is important to appreciate that a driven turbulent flow may experience,
over time, significant deviations from its average scaling laws, and that this may
be the explanation for
observed variations of the stellar IMF from place to place much in excess of the Poisson
variance related to the statistical sample size.  The scaling laws were understood 
phenomenologically by Kolmogorov as due to a scale independent 
energy dissipation rate, arising from an efficient energy cascade from large 
to small scales in turbulent flows. This transfer from large to small scales takes 
approximately a dynamical time of the outer scale. Therefore, in a driven flow, 
any variations of the energy injection rate on a time-scale of order the dynamical 
time causes a ``bump'' in inertial-range scaling laws that has to propagate down 
the turbulent cascade, until it reaches the small viscous scales after approximately 
a dynamical time of the outer scale. Because the typical lifetime of star-forming 
regions is comparable to this dynamical time (and star formation starts immediately 
when a molecular cloud is assembled), the turbulence can hardly be considered 
relaxed, and large variations of the IMF from place to place should be expected. 
These variations should not be interpreted as the lack of a universal process of star 
formation, but rather as the evidence of both its turbulent origin and its short lifetime.

\begin{figure}[ht]
\centerline{
\epsfxsize=8.6cm \epsfbox{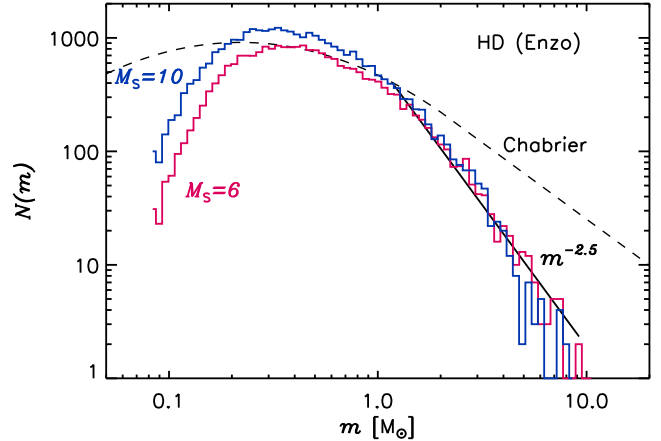}
}
\caption[]{Mass distributions of gravitationally unstable cores from the 
Enzo HD runs with $M_{\rm S}=6$ and $M_{\rm S}=10$, $f=8$\% and assuming a 
mean gas density of $10^4$~cm$^{-3}$ and a box size of 6~pc. Each mass 
distribution contains unstable cores from two snapshots. The 
Chabrier (2003) IMF (dashed line) and the power law predicted by the turbulent
fragmentation model (solid line) are also plotted.
}
\label{f13}
\end{figure}

\subsection{Previous Results}

Ballesteros-Paredes et al. (2006) argue that the fragmentation model 
of Padoan and Nordlund (2002) is in contradiction with their numerical results,
based on TVD and SPH simulations without magnetic fields. They conclude that the core mass 
distribution depends on the rms Mach number, but fail to point
out that the Padoan and Nordlund model contains such a Mach number dependence, 
with the peak of the mass distribution shifting to lower masses as the 
Mach number increases, in agreement with the numerical results in 
Ballesteros-Paredes et al. (2006). In the Padoan and Nordlund model, the 
slope of the mass distribution for masses above the peak is independent 
of the Mach number, also in agreement with the results of Ballesteros-Paredes
et al. (2002) based on the TVD simulations (see their Figure~4), but in
contradiction with their SPH simulations (see their Figure~5). 

Figure~\ref{f12} compares the power spectrum from the Stagger-Code HD run with 
two TVD and two SPH power spectra from Ballesteros-Paredes et al. (2006), 
for Mach numbers 3 and 6. The inertial range in both the TVD and SPH cases 
is not very extended, due to the low numerical resolution. The TVD code gives
a slope of $\beta\approx 2.2$, the same value found in the Zeus run, for both 
Mach numbers. The extent of the inertial range in the TVD run is also comparable 
to the Zeus result at the same resolution (not shown). The power spectra of the SPH 
runs are instead much steeper, and their slope increases with decreasing Mach number,
$\beta \approx 2.7$ for $M_{\rm S}\approx 6$ and $\beta \approx 2.9$ for
$M_{\rm S} \approx 3$. As shown by the TVD runs, the power spectrum should
not vary much with Mach number between $M_{\rm S}= 6$ and $M_{\rm S}=3$. For
lower Mach numbers, the power spectrum should become shallower, and converge
to a value of $\beta \approx 5/3$ for $M_{\rm S}<1$. The SPH power spectrum slope 
is therefore much too steep and its Mach number dependence unphysical.

In summary, it appears that the TVD runs of Ballesteros-Paredes et al. (2006)
are able to qualitatively reproduce the turbulent fragmentation process 
that we have tested in the present work with much larger numerical resolution, with three different 
grid-based codes and both with and without magnetic fields. The complete absence of an inertial range with a reasonable slope, 
or with a reasonable dependence of the slope on the Mach number, makes their 
SPH simulations totally inadequate for testing the turbulent fragmentation 
model, as the model relies on the scale-free nature of turbulent flows. 
Nevertheless, Ballesteros-Paredes et al. (2006) seem to base their conclusions 
partly on the SPH results, even if in contradiction with their own more 
robust TVD results that are confirmed here.

Ballesteros-Paredes et al. (2006) try to argue that the magnetic field
plays no role in the IMF, to justify the relevance of their simulations
without magnetic fields. However, we have shown here that the HD
regime produces much steeper mass distributions than the MHD regime.
The HD regime is therefore a poor choice of physical parameters, if the aim is to
extract the power law mass distribution from simulations with a very
limited numerical resolution, as it is much more variance-dominated
than the MHD regime. As explained above, it is not surprising that the very limited tails of
their power laws may possibly appear more Lognormal than straight.

To address directly the issue of the Mach number dependence of the mass
distribution, raised by Balesteros-Paredes et al. (2006), we plot in
Figure~\ref{f13} the mass distributions from the Enzo HD runs with $M_{\rm S}=6$ 
and $M_{\rm S}=10$ and with $f=8$\%. These mass distributions are computed assuming a 
mean gas density of $10^4$~cm$^{-3}$ and a box size of 6~pc. Each distribution
contains cores from two snapshots. The Figure shows that the power law
part of the mass distribution, above 1-2~M$_{\odot}$, is independent of the
Mach number and matches the prediction of the turbulent fragmentation model,
that is $k^{-2.5}$ for the power spectrum slope $\beta=1.9$ of the HD Enzo
runs.

\section{Conclusions}

We have used large numerical simulations of supersonic MHD and HD turbulence 
to test the turbulent fragmentation model of Padoan and Nordlund (2002).
The model predicts a power law distribution for large masses, related 
to the turbulence energy power spectrum slope, and the shock jump conditions. 
This power law mass distribution is confirmed by the numerical experiments. 
The model also predicts that the HD regime should yield a much steeper mass
distribution of unstable cores than the MHD regime, which is confirmed
by the simulations. This feature of the fragmentation model
is very important because it shows that even rather weak magnetic fields
(super-Alfv\'{e}nic turbulence) can be crucial in setting the initial conditions
for the process of star formation and in shaping the stellar IMF. 

While present-day star formation takes place probably always in the MHD regime, 
star formation at very high redshift may well occur in the HD regime, both 
because the field strength is still low and because the value of the critical 
magnetic field strength that defines the HD regime is larger at higher temperatures.
The IMF of the earliest population II stars, and perhaps the latest population
III stars as well, may be formed in turbulent environments in this HD regime, 
resulting in an IMF narrowly peaked around a mass of order 10~M$_{\odot}$.
The effect of such a peculiar IMF of early stellar populations on the ionization 
history and metallicity evolution of the universe should be investigated.

Numerical simulations can quantitatively account for the fundamental role of the 
turbulence in setting the initial conditions for the process of star formation 
only if they can generate an inertial range of turbulence,
which requires both low numerical diffusivity and large numerical resolution.
Furthermore, to model present-day star formation that occurs in the MHD 
regime, the magnetic field cannot be neglected, {\it even if the turbulence is
assumed to be super-Alfv\'{e}nic}. SPH simulations of large 
scale star formation to date fail in all three fronts: numerical diffusivity, 
numerical resolution and presence of magnetic fields. This should cast 
serious doubts on the value of comparing predictions based on
SPH simulations with observational data (see also Agertz et al. 2006).

Finally, the mass distribution of unstable cores found in the MHD simulations
is indistinguishable from the Chabrier stellar IMF (Chabrier 2003) and in
qualitative agreement with the less well determined mass distributions of
prestellar cores selected from dust emission or molecular line observations.
Such a coincidence may indicate that gravitational fragmentation, competitive 
accretion or merging, all absent in these turbulence simulations, may not play 
a major role in the origin of the stellar IMF, a fascinating idea to be tested
with the next generation of high resolution simulations of self-gravitating
turbulence.

\acknowledgements

This research was partially supported by a NASA ATP grant NNG056601G,
by an NSF grant AST-0507768, and by a NRAC allocation MCA098020S.
We utilized computing resources provided by the San Diego Supercomputer 
Center, by the National Center for Supercomputing Applications and by NASA 
High End Computing Program.

\bibliographystyle{apj}
%\bibliography{apj-jour,MC,padoan,bd,enzo_tech2004.bib}

\end{document}